\documentclass[debug,overfull]{epl}
\title{Density Wave and Supersolid Phases of Correlated Bosons in
an optical lattice}
\author{Dmitry L. Kovrizhin\inst{1,2} \thanks{ Email : [dima, venkat,
 subhasis]@mpipks-dresden.mpg.de}
\and G. Venketeswara Pai\inst{1}
\and Subhasis Sinha\inst{1}}
\shortauthor{D. L. Kovrizhin \etal}
\shorttitle{Density wave and supersolid phases \dots}
\institute{
\inst{1} Max-Planck-Institut f\"{u}r Physik komplexer Systeme,
N\"{o}thnitzer Stra\ss e 38, 01187 Dresden, Germany\\
\inst{2} RRC Kurchatov Institute, Kurchatov Sq. 1, 123182 Moscow, Russia
}

\begin{document}

\maketitle

\begin{abstract}
Motivated by the recent experiment on the Bose-Einstein condensation of
$^{52}$Cr atoms with long-range dipolar interactions (Werner J. et al.,
Phys. Rev. Lett., 94 (2005) 183201),
we consider a system of bosons with repulsive
nearest and next-nearest neighbor interactions in an optical lattice. 
The ground state phase diagram, calculated using the
Gutzwiller ansatz, shows, apart from the superfluid (SF) and the Mott insulator
(MI), two modulated phases, \textit{i.e.}, the charge density wave (CDW) and
the supersolid (SS). Excitation spectra are also calculated which show a gap in
the insulators, gapless, phonon mode in the superfluid and the supersolid,
and a mode softening of superfluid excitations in the vicinity of the
modulated phases. We discuss the possibility of observing these phases in
cold dipolar atoms and propose experiments to detect them.
\end{abstract}

The recent observation \cite{bloch} of quantum phase transition of
interacting bosons in an optical lattice has given a new impetus to the
study of correlated bosons. The possibility of manipulating both the
interactions among the constituents as well as disorder in the system with
extreme control, make them an ideal object over conventional solid state
systems to study many-body effects. Recent experimental success in obtaining
a dipolar condensate of $^{52}$Cr atoms \cite{crexpt} opens up new
directions in the study of lattice bosons with long-range interactions \cite%
{goral}. This also leads to the possibility of achieving new phases such
as insulating charge density wave and the supersolid that has both superfluid
and crystalline properties \cite{sstheory}. 
The recent experimental observation of
nonclassical moment of inertia of solid $^{4}$He indicates a possible
signature of SS \cite{ssexpt}, although the existence of the SS is still a
controversial issue, and we believe, the dipolar condensate in optical
lattice can provide a testing ground for it.

A collection of interacting bosons on a lattice is described by the
Bose-Hubbard model \cite{fisher} which captures the physics arising from the
competition between kinetic energy of bosons and their on-site interaction.
However, long-range interactions among the dipolar bosons at different 
lattice sites
make it necessary to include nearest-neighbor interaction in the
Hamiltonian. The resulting extended Bose-Hubbard model (eBHM) Hamitonian reads: 
\begin{equation}
\hat{H}=-t\sum_{i,\delta }\hat{a}_{i}^{+}\hat{a}_{i+\delta }+\frac{U}{2}%
\sum_{i}\hat{n}_{i}\left( \hat{n}_{i}-1\right) +V\sum_{i,\delta }\hat{n}_{i}%
\hat{n}_{i+\delta }.  \label{1}
\end{equation}%
Here $\hat{a}_{i}^{+}$ creates a boson at site $i$, $\hat{n}_{i}=\hat{a}%
_{i}^{+}\hat{a}_{i}$\ is the boson number operator, $t$, $U$, $V$, are the
nearest-neighbor hopping, the on-site interaction and the nearest-neighbor
interaction (NNI) respectively, and $\delta $ represents the nearest
neighbors of site $i$. This model in a high density limit, where it
corresponds to a quantum phase model (QPM), has also been studied in the 
context of a wide
variety of other problems such as Josephson junction arrays, granular
superconductors, liquid $^{4}$He in Vycor \textit{etc.} \cite{fazio,otterlo}.
For small $t$, the role of NNI is to stabilize an insulating state,
similar to the CDW phase in solids, where the number density alternates at
every lattice site between two successive integers. In the case of soft-core
bosons, it also brings in a new phase where both diagonal crystalline order
and off-diagonal superfluid long-range order coexist and hence called the
supersolid \cite{sstheory}; in this phase the number density shows a
modulation, while still retaining a nonvanishing SF fraction.
The model in Eq. (\ref{1}) without the NNI has been extensively studied
previously using different methods \cite{bhm, krauth, fisher}, and the
mean-field results were found to be in good agreement with the quantum
Monte-Carlo (QMC) results \cite{bhm}, though the critical behavior is beyond
the description of the former. However, one can expect that with the increasing
dimensionality $d$ of the system, larger filling factors, and the longer
range of the interactions, mean-field results will progressively become
accurate. In this letter we study the eBHM using Gutzwiller mean field theory \cite%
{jaksch} to obtain the ground state phase diagram. More importantly, we
extend this approach, using the time dependent variational principle, to
extract the excitation spectrum of the system. This method has the added
advantage of generalizing to the nonuniform systems, for example in a
confining trap, which is always present in experiments. We discuss the
possibility of observing these novel phases in cold atoms and propose ways
to detect them.

In order to find the ground state of the Hamiltonian (\ref{1}),
we use the variational minimization of 
$\left\langle \Psi \right\vert \hat{H}-\mu \hat{N}\left\vert \Psi
\right\rangle $ with a Gutzwiller wavefunction $\left\vert \Psi
\right\rangle ={\textstyle\prod\nolimits_{i}}\sum_{n}f_{n}^{(i)}\left\vert
n_{i}\right\rangle $ where $\left\vert n_{i}\right\rangle $ is the Fock
state with $n$ particles at site $i$, $\mu $ is the chemical potential, and $%
\hat{N}=\sum_{i}\hat{n}_{i}$ is the total particle number operator.
Minimization with respect to the variational parameters $f_{n}^{(i)}$ gives
the ground state, which could be a MI, SF, SS, or a CDW. First we consider the
case when $Vd \leq U/2$. In Fig. 1, we present
the phase diagram for $Vd/U=0.4$ \cite{dimensionality}.
For small enough $t$, incompressible MI and CDW phases are
formed. The MI phase with integer $n_{0}$ number of particles per site
(denoted by MI($n_{0}$)) is characterized by $f_{n}^{(i)}=\delta _{n,n_{0}}$
and a vanishing condensate order parameter $\phi _{i}=\sum_{\delta
}\left\langle \Psi \right\vert \hat{a}_{i+\delta }\left\vert \Psi
\right\rangle $. The CDW phase has a two sublattice periodic structure such
as $n_{0},n_{0}-1$ particles at the consecutive sites (denoted by CDW($%
n_{0}/2$)) and zero $\phi _{i}$. The SF and the SS phases have non-vanishing 
$\phi $; in the former it is uniform and in the latter it has a
two-sublattice modulation. At small values of $\mu $ and $t=0$, we have a
CDW(1/2) with $\left\vert 1,0,1,0\right\rangle $ particle density
modulation. Upon further increase of $\mu $, MI(1) phase becomes 
energetically favourable. In general, at $t=0$, we have
transitions between CDW($n_{0}/2$) to MI($n_{0}$) at $\mu
= U(n_{0}-1)+2Vdn_{0}$ and then to CDW($n_{0}/2+1$) at $\mu 
= Un_{0}+2Vdn_{0}$.
As we lower the value of $V$, the area of the CDW lobe reduces 
continuously and so is
the transition point at $t=0$ between the CDW and the MI phases; the CDW
lobe vanishes completely for $V=0$, thereby reducing to the phase diagram as
in Ref.~\cite{fisher}. Uniform, homogeneous SF phase appears for
sufficiently large values of $t$. In between the CDW and the SF, a
supersolid (SS) phase is formed, which has a density modulation as well as
nonvanishing~$\phi $. Here we must stress that a \textit{finite} onsite
interaction is needed, in addition to the NNI, for the existence of 
checkerboard SS phase which is unstable in the
hard-core limit of eBHM \cite{batrouni}. As $t$ increases, the system
undergoes a second-order transition from CDW(1/2) to a SS for larger values
of $\mu $ (mostly for particle-doping away from half-filling) and a first
order transition to SF at smaller values of $\mu $ (for hole-doping away
from half-filling). A multicritical point exists at $\mu/U =0.132$ and 
$td/U=0.137$
where the two second order transition lines between SS-SF and CDW-SS meet
the first order line between CDW-SF (see Fig. 1). In Fig. 2a, we plot the
contours of fixed particle density as a function of $\mu $ and $t$ for $%
Vd/U=0.4$. We find the insulating lobes of fixed average commensurate
densities $n_{0}=0.5$, and $n_{0}=1$ \textit{etc.}, and SF and SS phases
with incommensurate densities (in general). 
As shown in Fig. 2b and Fig. 2c, we observe at a given $\mu$, the density
and $\phi$ change continuously across CDW-SS and SS-SF 
transitions indicating a continuous second order transition, whereas a 
jump in the above quantities signals a first order transition between 
CDW-SF.  

When $Vd > U/2$, the MI lobes vanish and the new CDW phases
such as CDW(1) with a particle density modulation $\left| \dots, 2, 0,
\dots \right>$ appear \cite{pandit}. At $t=0$, the CDW(1/2) transforms 
into CDW(1)
when $\mu > U$, which in turn transforms into CDW(3/2) when
$\mu > 2U$. In general a $|\dots,n_0,0,\dots>$ CDW state is
formed within a region $(n_0 - 1) U < \mu < n_0 U$.
With increasing $t$, the supersolid phase is formed around
the CDW lobes and at still larger values of $t$, a homogeneous superfluid 
phase appears.

In 1D due to enhanced quantum fluctuations, our results
differ significantly from those of the density matrix renormalization group
where no SS phase was found \cite{white}. In fact, the numerical studies
of 1D BHM show that the nature of the Mott-SF phase boundary itself
is quite different from that obtained from the Gutzwiller method.
In 2D previous QMC studies for the hard-core ($U = \infty$)
bosons found unstable checkerboard SS phase 
which phase separates between CDW and SF \cite{batrouni}.
Preliminary results for the soft core bosons indicate the region of
SS phase with negative compressibility \cite{scalettar}.
However, a previous QMC study of the related quantum phase model
by van Otterlo {\it et al.} \cite{otterlo} came to the conclusion
that a finite $U$ (soft core) stabilizes the checkerboard SS phase.
Within the Gutzwiller approach, we find a stable SS phase over a large
region of particle doping, and a relatively smaller region of
SS upon hole doping the CDW, and a first order transition between the CDW
and the SF below the multicritical point (region of hole doping)
indicating possible phase separation.
A recent QMC study {\it indeed} shows that a stable SS phase exists 
for soft-core bosons with sufficiently large values of $V/U$ and upon
particle-doping away from half-filling \cite{troyer}; however, the SS
obtained by hole doping turns out to be unstable with respect to
phase separation. The CDW-SF transition is indeed found to be first order
as expected.
The region of SS phase
reduces significantly with reducing $V$, and we expect that it will be
destroyed at small enough $V$ due to quantum fluctuations. Increasing the
dimensionality to three and/or increasing the filling factor 
will reduce the fluctuations, thus stabilizing the
SS further.

\begin{figure}
\twofigures[scale=0.5]{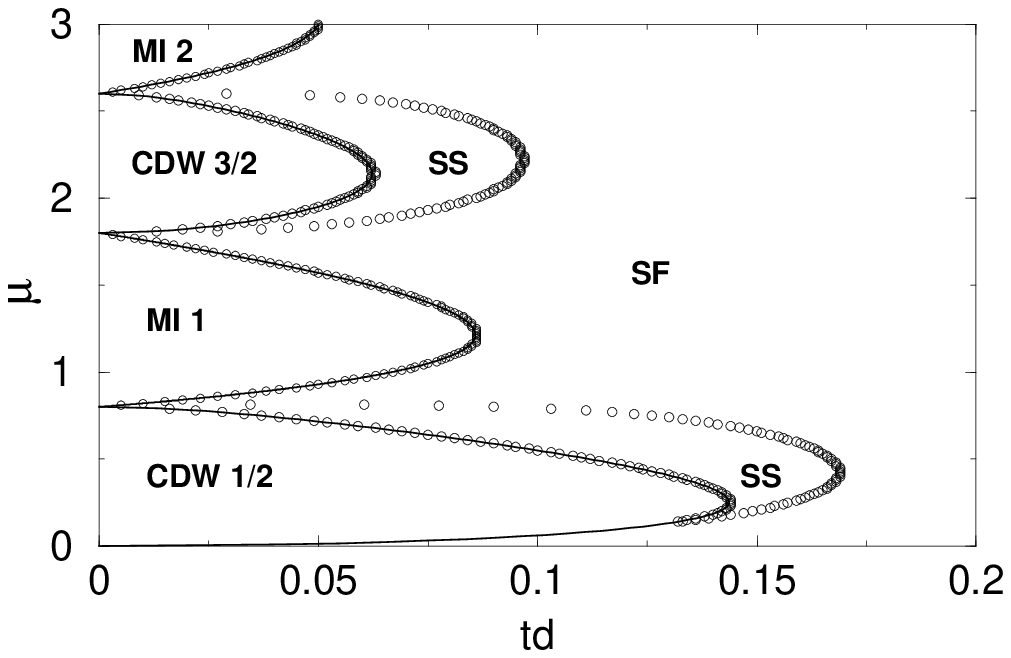}{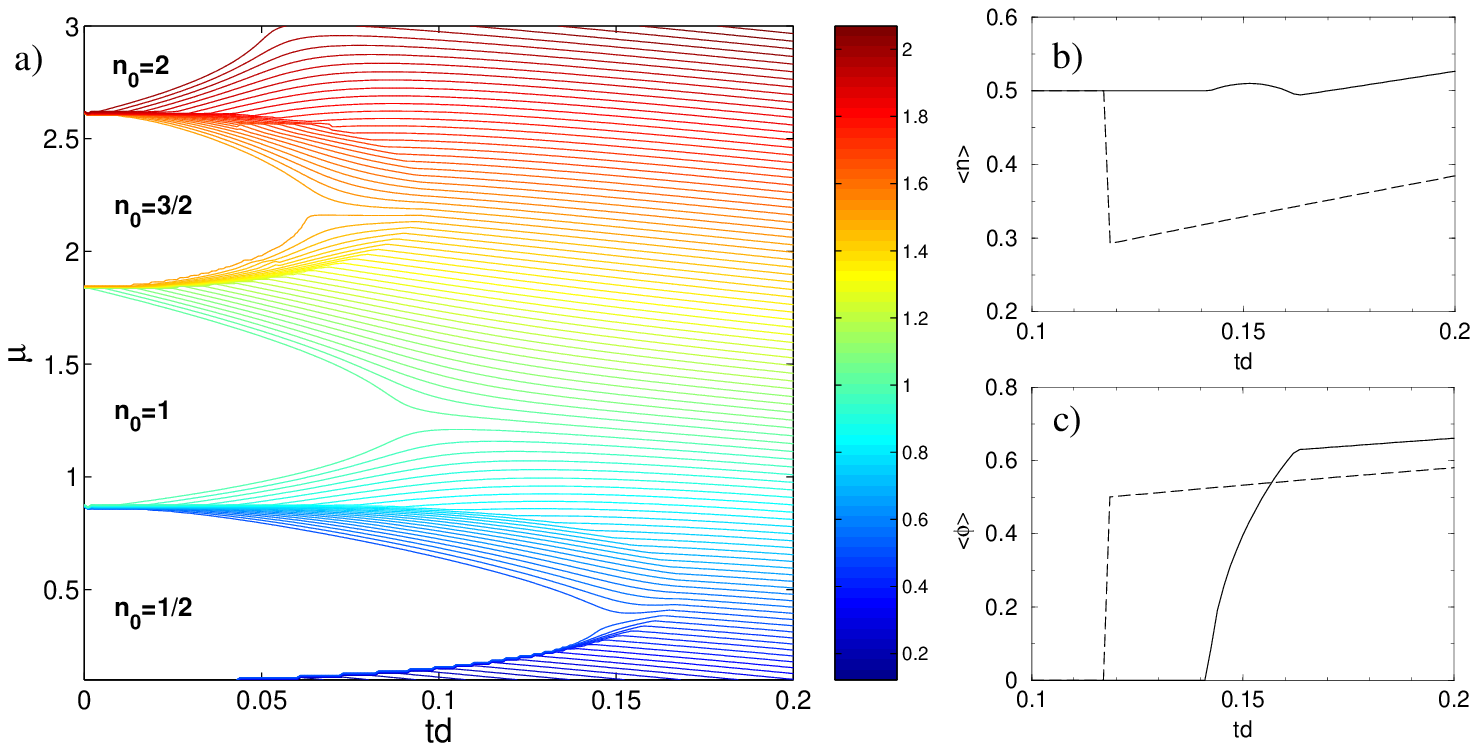}
\caption{Phase diagram of the eBHM for $Vd=0.4$ (energies are in units of 
U). Open circles denote the
numerical results and the solid lines represent the analytical results.}
\label{fig:1}
\caption{a) Contour plot of fixed particle density for $Vd=0.4$ (energies 
are in units of U). b)
Variation of the average particle density for a fixed $\protect\mu $ and for
$Vd=0.4$. Solid line: CDW to SS to SF transition at $\protect\mu =0.3$.
Dashed line: CDW to SF transition at $\protect\mu =0.1$. c)~Variation of $%
\protect\phi $ for the same parameters.}
\label{fig:2}
\end{figure}

The excitation spectrum of the system can be obtained from the dynamical Gutzwiller
approach with the variational parameters $f_{n}^{\left( i\right) }$ being
time dependent. Minimization of the effective action $\left\langle \Psi
\right\vert i\frac{\partial }{\partial t}-\hat{H}+\mu \hat{N}\left\vert \Psi
\right\rangle $ gives the equations of motion for 
$f_{n}^{\left( i\right)}$:
\begin{eqnarray}
i\frac{\partial f_{n}^{\left( i\right)}}{\partial\tau} &=&\left[
\frac{U}{2}n\left( n-1\right)
-\mu n+Vn\rho ^{\left( i\right) }\right] f_{n}^{\left( i\right) } 
-t\left( \phi _{i}^{\ast }\sqrt{n+1}f_{n+1}^{\left( i\right) }+\phi
_{i}^{\ast }\sqrt{n}f_{n-1}^{\left( i\right) }\right)  \label{2}
\end{eqnarray}%
where $\rho ^{(i)}=\sum_{\delta ,n}n{|f_{n}^{(i+\delta )}|}^{2}$ and 
$\tau$ is time. The small
amplitude fluctuations $\delta f_{n}^{(i)}(t)$ around the ground state $\bar{%
f}_{n}^{(i)}$ give the excitation spectrum. In the MI ground state, $%
\delta f_{n}^{(i)}$ are connected only to $\delta f_{n\pm 1}^{(i)}$ in the
Eq. (\ref{2}). The low-lying excitations are creating particles (p) and
holes (h) in the ground state with a particle density $n_{0}$ which needs a
finite energy and the system is gapped. At $t=0$, these excitations have
energies $U n_{0} + 2dVn_{0} - \mu$ and $\mu - U(n_{0} - 1) - 2dVn_{0}$ 
respectively. For nonzero $t$, the dispersions of these excitations are: 
\begin{eqnarray} \label{mottex}
\omega _{ph}=\pm [-\epsilon /2+ U
(n_{0}-1/2) + 2dVn_{0} - \mu]+{[{(\epsilon /2)}^{2}-\epsilon
U(n_{0}+1/2)+U^{2}/4]}^{1/2} 
\end{eqnarray}
where $\epsilon
(\mathbf{k})=2t\sum_{l=1..d}\cos (k_{l})$. For $V=0$ this result agrees with
the excitation spectra of the MI phase of BHM obtained within the slave boson
approach \cite{stoof}. In general, the  CDW state has four 
low-lying excitations 
in the reduced Brillouin zone (BZ), corresponding to particle and hole 
excitations
in each of the A (with $n_{1}$ particles per site) and B (with $n_{2}$
particles per site) sublattices.
At $t=0$, the energies of particle (hole) excitations of sublattice A
are $E_{A}^{p}=Un_{1}+2dVn_{2}-\mu ,\
E_{A}^{h}=-U(n_{1} - 1)-2dVn_{2} +\mu$ 
,and those of sublattice B are $E_{B}^{p}=Un_{2}+2dVn_{1}-\mu $,
and $E_{B}^{h}=-U(n_{2} - 1)-2dVn_{1}+\mu $ respectively which, as in the 
case of MI, pick up a dispersion at nonzero values of $t$. 
The number of particles on sublattice B is $n_{2} = n_{1} - 1$, for 
$Vd < U/2$ and $n_{2} = 0$, for $Vd > U/2$. 
Clearly, the hole excitation in sublattice B is unphysical for $n_{2} = 0$.
For $Vd > U/2$, there is a qualitative change
in the excitation spectrum. The excitation corresponding to a particle
addition on A becomes energetically favourable compared to a particle
addition on B. At finite $t$ the excitation spectrum $\omega(\mathbf{k})$ 
of the insulating 
states can be obtained analytically by solving the algebraic equation:
\begin{equation} \label{exc}
{\tilde E_A}^p{\tilde E_A}^h{\tilde E_B}^p{\tilde E_B}^h
-{\epsilon}^{2}(\mathbf k) [(n_{1} +
1) {\tilde E_{A}}^{h} + n_{1} {\tilde E_{A}}^{p}]
[(n_{2} + 1) {\tilde E_{B}}^{h} + n_{2}{\tilde E_{B}}^{p}] = 0,
\end{equation}
where $\tilde{E}_{A(B)}^{p(h)}=E_{A(B)}^{p(h)}-\omega$. 
Three low-lying excitations for the CDW(1/2) are shown in Fig. 3a.
The excitations of the MI
state can also be obtained from the Eq. (\ref{mottex}) by
substituting $n_{1} = n_{2}$. At the phase boundaries of the
insulating states excitations become gapless and the analytical form of 
the phase boundaries can also be obtained from Eq. (\ref{exc}) by 
substituting $\omega=0$ and $\mathbf {k}=0$ . 
These are represented by solid lines along with the
numerically obtained results in Fig. 1.

\begin{figure}
\twofigures[width=6cm]{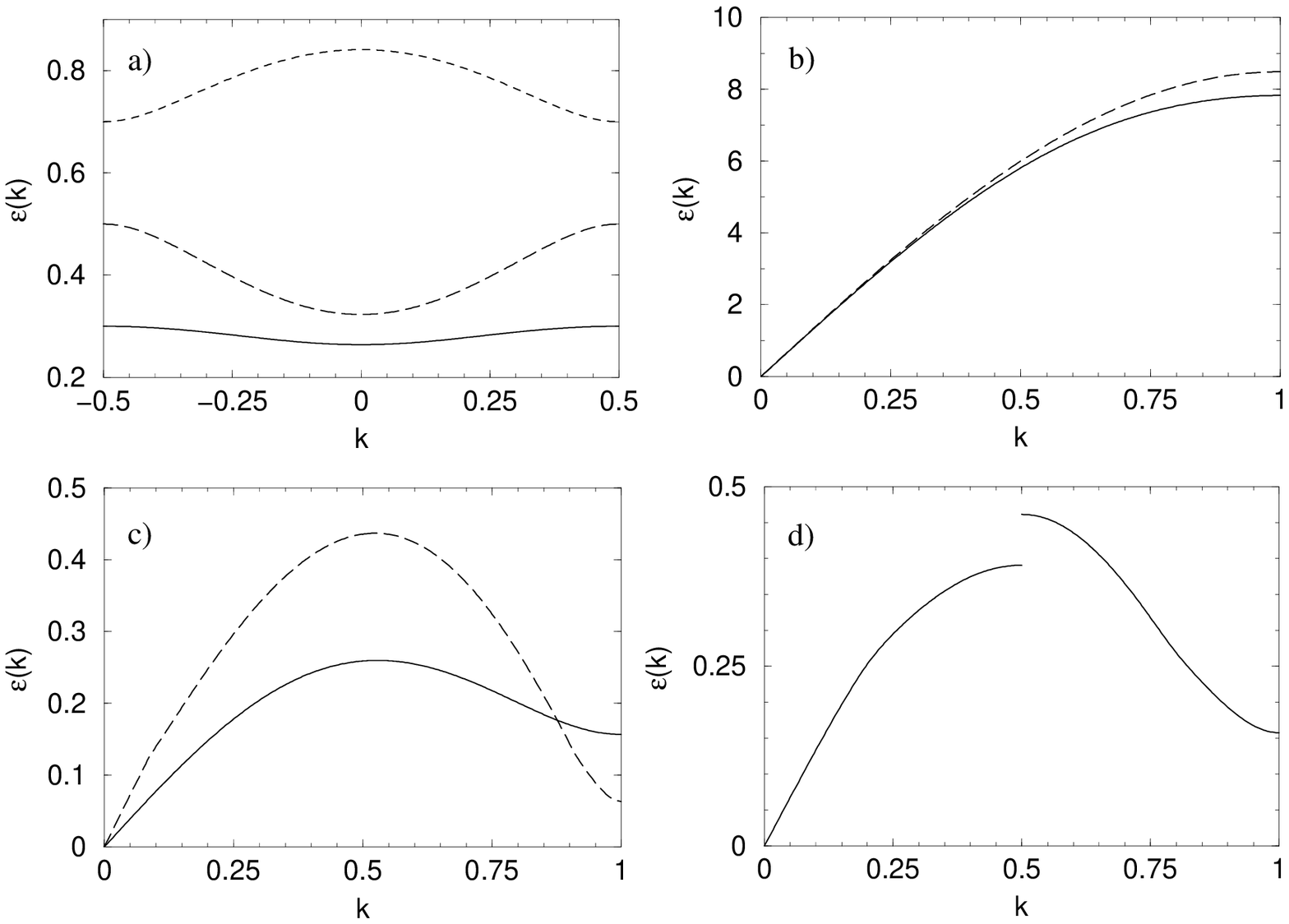}{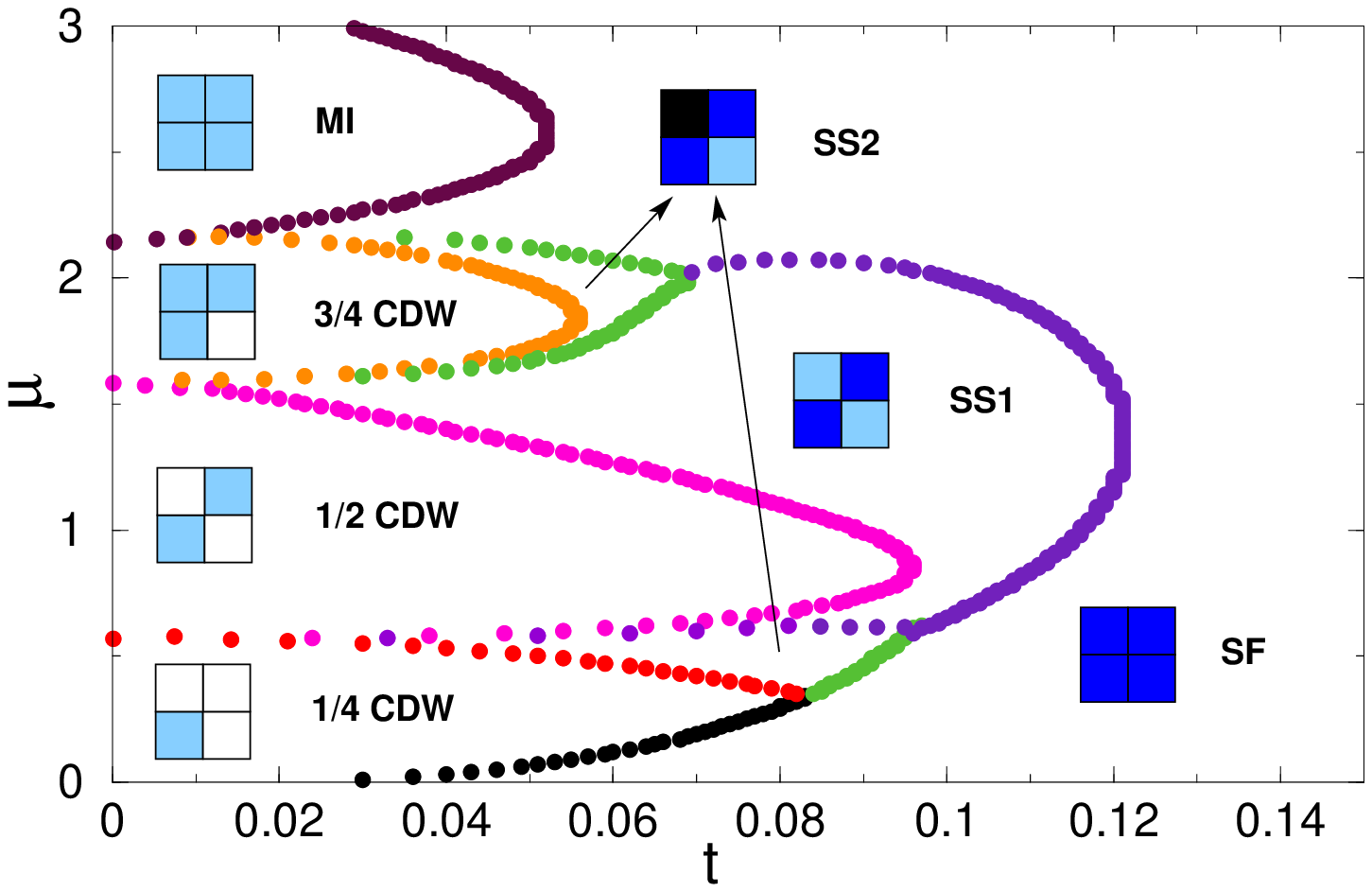}
\caption{Excitation spectra for $Vd=0.4$ along $k_{x}/\protect\pi =k_{y}/%
\protect\pi =k$. a) Excitation spectra of CDW ($td=0.08,\protect\mu =0.3$),
solid line: hole excitation on sublattice A, dashed line: particle
excitation on sublattice B, short-dashed line: particle excitation on A b)
Spectrum in the deep SF regime (solid line with $td=2,\protect\mu =0.4$),
dashed line is the excitation spectrum of the effective DNLS for the same
parameters. c) solid line: the SF excitation spectrum near SS boundary ($%
td=0.17,\protect\mu =0.4$), dashed line: the SF spectrum close to CDW phase (%
$t=0.12,\protect\mu =0.05$), notice the mode softening near the BZ boundary.
d)~SS excitation spectrum for $td=0.16,\protect\mu =0.4$ (energies are 
in units of U).}
\label{fig:3}
\caption{Phase diagram of the eBHM in 2D with $V=0.4$ and $V^{\prime}=0.4/(2%
\protect\sqrt{2})$ (energies are in units of U).}
\label{fig:4}
\end{figure}

In the SF phase, with the ground state values $\bar{f}_{n}^{\left( i\right)}$,
one can also get the spectra of the excitations from Eq. (\ref{2}).
These quasiparticles, with mixed particle-hole character, are gapless and
have phonon-like linear dispersion (Fig. 3b). Deep in the SF (for large
values of $t$), however, $\bar{f}_{n}^{\left( i\right) }$ obey the
Poissonian distribution $\bar{f}_{n}^{\left( i \right)} = \phi_i^n
\exp{\left(-{\left|\phi_i \right|}^2/2\right)}/\sqrt{n!}$
and Eq. (\ref{2}) reduces to a discrete nonlinear
Schr\"{o}dinger equation (DNLS):
\begin{eqnarray}
i \frac{\partial \phi_i}{\partial{\tau}} = -t\sum_{\delta} \phi_{i + 
\delta} + U {\left| \phi_i \right|}^2 \phi_i
+V \sum_\delta {\left| \phi_{i+\delta}
\right|}^2 \phi_i - \mu \phi_{i}.
\end{eqnarray}
In this limit the excitation spectrum can
be obtained analytically (dashed line in Fig. 3b) with the sound velocity $%
[2td(U+2Vd){|\phi |}^{2}]^{1/2}$ and it matches very well with the numerical
result, obtained from Eq. (\ref{2}). However, as one moves towards the
insulating phases, $|\phi|^{2}$ deviates significantly from 
$\rho$, and
the description in terms of the DNLS becomes progressively inaccurate. As we
approach the MI boundary, $\phi $ reduces and finally vanishes at the
transition point signaling a continuous transition at which a gap opens up
for the particle and the hole excitations. This behavior has to be contrasted
with the approach towards the modulated phases (SF-CDW and SF-SS) where a
mode softening in the SF spectrum at $\mathbf{k}=(\pi ,\pi )$ implies the
instability of the homogeneous SF phase and appearance of states with broken
translational symmetry \cite{nozieres} (Fig. 3c). Once the SS boundary is
crossed, we observe a gapless, phonon-like mode in the reduced BZ due to
nonvanishing SF fraction, as well as higher energy gapped mode as shown in
Fig. 3d. The appearance of the gap can be understood as due to the doubling
of the unit cell and the Bragg reflection at $\mathbf{k}=(\pi /2,\pi /2)$.

Let us now discuss the possibility of observing these phases experimentally.
The novel, translational symmetry broken phases such as SS and CDW appear
because of the longer range interactions between bosons. There are many ways
which we could use to achieve bosonic systems with such interactions.
One possibility is to use Rydberg atoms; however, they have short life-times
that make them difficult candidates to observe these phenomena. The most
promising candidate is a dipolar condensate of $^{52}$Cr atoms which have a
large magnetic dipole moment \cite{crexpt}. A spin-polarized dipolar system,
to the leading order, has a power-law tail with $V_{ij}\sim \mu _{d}^{2}/{%
\left\vert {\bf r}_{i}-{\bf r}_{j}\right\vert }^{3}$ where $\mu _{d}$ is the dipole
moment and $i,j$ are two lattice sites. The on-site interaction $U$ can be
tuned by Feshbach resonance and $t$ by the intensity of the laser beams
which form the optical lattice. Polarization of the dipoles can be
manipulated by external magnetic field which can produce repulsive
interaction between the atoms. Once such a condensate is formed in an
optical lattice and interactions are tuned by external sources, the phases
such as the CDW and the SS could be observed in experiments similar to Ref. 
\cite{bloch}. Both CDW and SS phases are characterized by modulation of
the particle density in real space imaging (since both of them have a
crystalline order). In addition the CDW will have a gap in the excitation
spectrum. The SS phase will show phase coherence in the time of flight
experiments. In particular, we find that momentum distribution 
$n(\mathbf{k})$ in a $N\times
N$ lattice sites is given by: 
\begin{eqnarray}
n(\mathbf{k})&=&(\rho_{A}+\rho_{B})/2+({|\phi _{A}|}^2
+{|\phi _{B}|}^{2})(\alpha _{x}\alpha _{y}-1)/2 
+\alpha _{x}\alpha _{y}({
|\phi _{A}|}^{2}\cos (k_{x}+k_{y})\nonumber \\
& + &{|\phi _{B}|}^{2}\cos
(k_{x}-k_{y}))/2
+\phi _{A}\phi _{B}\alpha _{x}\alpha _{y}(\cos k_{x}+\cos k_{y})
\end{eqnarray}
where $\alpha _{x/y}=(2/N)\sin ^{2}(k_{x/y}N/2)/\sin ^{2}(k_{x/y})$.
In the SF, $\rho_{A}=\rho_{B}$ and $\phi _{A}=\phi _{B}$. We see that 
contrary to
the SF case, where the momentum distribution has a sharp central peak, the
SS is characterized by additional peaks at $(\pm \pi ,\pm \pi )$, (small
compared to the central peak) characteristic of the ordering wave-vector of
the checkerboard phase. We have checked that the presence of a confining
harmonic trap does not destroy the CDW and the SS phases, but leads to
coexistence of different phases \cite{confined}. Time dependent Gutzwiller
theory can be applied to calculate the excitation spectra in trapped systems
as well \cite{dima}.

Finally, we consider the effect of next-nearest neighbor interaction (NNNI),\\
$V^{\prime }\sum_{i,\delta{'}}n_{i}n_{i+\delta^{'}}$, on the phase diagram,
where  $\delta^{'}$ is next-nearest neighbor of site $i$. In 2D for 
$V^{\prime} = V/(2\sqrt{2})$ (relevant for dipolar interaction),
we plot the phase diagram in Fig. 4. Checkerboard CDW and SS
are formed for $V>2V^{\prime }$.
In addition new phases appear, such as: CDW with one particle (CDW(1/4)) and
three particles (CDW(3/4)) in $2\times 2$ sublattice and SS with three
sublattice modulation (SS2) (see Fig. 4). Similar phases can also be
obtained by mapping hard-core bosons to Heisenberg spin model with NNNI \cite%
{scalettar,frey2}, although the phase diagram is different from the
soft-core boson model. In real dipolar gases, the long-range nature of
the interaction (which is truncated to NNNI in this work) as well as
anisotropy (which is also neglected here) could be important. However, as
can be seen from Fig. 4., inclusion of further interactions introduces
CDW states of different filling fractions and CDW and SS of different ordering
wavevectors; we do not expect fundamentally new phases to appear due to
the long-range interaction. It is also clear that striped CDW or SS
phases with $(\pi, 0)$ ordering will be absent in real dipolar gases
since $2V^{\prime} < V$. Even for realistic dipolar interactions, with the 
anisotropy (e.g. an attractive potential along the $z$-axis when the dipoles
are polarized in the $x-y$ plane) included, we expect, for example, CDW and
SS phases with $(\pi, \pi, 0)$ wavevector, if a collapsed
phase is avoided by having a large on-site repulsion that can be achieved
using a Feshbach resonance or alternatively, by having pancake-shaped
condensates. This, however, needs further study. Preliminary results
indicate that this indeed is happening.

In conclusion, we have studied the lattice bosons with NNI and presented
their ground state phase diagram and the low-lying excitations in the four
phases, by a time-dependent generalization of the Gutzwiller ansatz. In
particular, the SF excitations exhibit a mode softening at the BZ boundary
as phases with broken translational symmetry are approached.
We also studied the effect of NNNI
on the phase diagram of the lattice bosons. We have proposed that such phases
could be seen in dipolar condensates and discussed a number of ways to
detect them. Recent success in achieving dipolar condensate of $^{52}$Cr
atoms and manipulating the interactions \cite{crexpt}, 
we believe, could lead to the detection of
these interesting phases, particularly the SS, as well as creation of
frustrated lattice bosons \cite{murthy,santos} and the study of 
many associated interesting phenomena yet to be explored.

\acknowledgments
We thank I. Bloch, P. Fulde, M. Lewenstein, P. Nozi\`{e}res, F. Pistolesi, 
L. Santos, and K. Sheshadri for many valuable discussions. One of us (DLK)
was supported by INTAS-2001-2344 and NWO grants.


\begin{thebibliography}{0}
\bibitem{dimensionality} 
Our mean-field results have been extrapolated to $d$%
-dimensionality by rescaling parameters $V$ and $t$ by $1/d$.
\end{thebibliography}
\end{document}